\documentclass{ckm}                 
\newcommand{\spipi}{{\cal S}_{\pi\pi}}
\newcommand{\apipi}{{\cal A}_{\pi\pi}}
\newcommand{\DCRP}{\Delta{C}_{\rp}}
\newcommand{\DSRP}{\Delta{S}_{\rp}}
\newcommand{\Bzbarrp}{{\Bzbar_{\rp}}}
\newcommand{\Bzrp}{\Bz_{\rp}}
\newcommand{\rto}{\rightarrow}
\newcommand{\ArpCP}{A^{\rp}_{CP}}
\newcommand{\Crp}{C_{\rp}}
\newcommand{\Srp}{S_{\rp}}
\newcommand{\Crh}{C_{\rh}}
\newcommand{\Srh}{S_{\rh}}
\newcommand{\ArhCP}{A^{\rh}_{CP}}
\newcommand{\ArKCP}{A^{\rho K}_{CP}}
\newcommand{\dm}{\Delta m_d}
\newcommand{\dt}{\Delta t}
\newcommand{\Bz}{B^0}
\newcommand{\Bzbar}{\bar{B}^0}
\newcommand{\Bbar}{\bar{B}}
\newcommand{\rp}{\rho\pi}
\newcommand{\rh}{\rho h}
\newcommand{\rppm}{\rho^+\pi^-}
\newcommand{\rmpp}{\rho^-\pi^+}

\newcommand{\Tbz}{\tau_{B^0}}

\newcommand{\lampipi}{\lambda_{\pi\pi}}
\newcommand{\Apm}{A_{+-}}
\newcommand{\Amp}{A_{-+}}
\newcommand{\CrK}{C_{\rho{K}}}
\newcommand{\dCrK}{\Delta\CrK}
\newcommand{\SrK}{S_{\rho{K}}}
\newcommand{\dSrK}{\Delta\SrK}
\newcommand{\sTb}{\sin{2\beta}}
\newcommand{\sTfO}{\sin{2\phi_1}}
\newcommand{\sTa}{\sin{2\alpha}}

\newcommand{\Btopippim}{\Bz\rto\pi^+\pi^-}
\newcommand{\Btorp}{\Bz\rto\rho\pi}
\newcommand{\cpipi}{{\cal C}_{\pi\pi}}
\newcommand{\fcp}{f_{CP}}
\newcommand{\ftag}{f_{\rm tag}}
\newcommand{\tcp}{t_{CP}}
\newcommand{\ttag}{t_{\rm tag}}
\newcommand{\qq}{q\bar{q}}
\newcommand{\dE}{\Delta{E}}
\newcommand{\Mbc}{M_{\rm bc}}
\newcommand{\EBcms}{E^{cms}_B}
\newcommand{\Ebeamcms}{E^{cms}_{beam}}
\newcommand{\pBcms}{p^{cms}_B}
\newcommand{\rpipi}{{\cal R}_{\pi\pi}}

\usepackage{amsmath}

\usepackage{txfonts}            

\confname{Workshop on the CKM Unitarity Triangle, IPPP Durham, April
  2003}

\title{Status and perspectives of $\sTa$ measurements}
\author{H Sagawa\addressmark{a}}

\address[a]{KEK, Tsukuba}

\begin{document}

\begin{abstract}
In the neutral $B$ meson system, it is possible to measure  
the CKM angle $\alpha$ using the decay mode $b\rto u\bar{u}d$ 
in the presence of pollution from gluonic $b\rto d$ penguin decays.
Here the recent status of the measurements of \textit{CP}-violating
asymmetry parameters using time-dependent analyses in
$\Btopippim$ and $\Btorp$ decays and 
the perspectives of a $\sTa$ measurement are presented.
\end{abstract}

\maketitle


\section{Introduction}



In 1973, Kobayashi and Maskawa (KM) proposed a model where \textit{CP} violation is
accommodated as an irreducible complex phase in the weak-interaction quark
mixing matrix~\cite{KM}.
Recent measurements of the \textit{CP}-violating asymmetry parameters $\sTb ~(=\sTfO)$
~\cite{bib:angle-def}
by the Belle~\cite{bib:sin2phi1-Belle} and BaBar~\cite{bib:sin2phi1-BaBar}
 Collaborations established \textit{CP} violation in the
neutral $B$ meson system.
Measurements of other \textit{CP}-violating asymmetry parameters provide important tests
of the KM model. 
Any mode with a contribution from $b \rto u\bar{u}d$ is a possible source of
measurement of the Cabibbo-Kobayashi-Maskawa (CKM) angle $\alpha ~( = \phi_2 )$.
Here the recent status of the measurements of \textit{CP}-violating asymmetry
parameters using time-dependent analyses in
$\Btopippim$ and $\Btorp$ decays~\cite{bib:CC} and 
the perspectives of a $\sTa$ measurement are presented.

\section{$\Bz\rto\pi^+\pi^-$ decays}

The $\Btopippim$ decay is one of the important modes for the measurement of
$\sin 2\alpha$.
The KM model predicts \textit{CP}-violating asymmetries in the time-dependent rates
for $\Bz$ and $\Bzbar$ decays to a common \textit{CP} eigenstate, $\fcp$.
In the decay chain $\Upsilon(4S)\rto\Bz\Bzbar\rto\fcp\ftag$, in which
one of the $B$ mesons decays at time $\tcp$ to $\fcp$ and the other decays at
time $\ttag$ to a final state $\ftag$ that distinguishes between
$\Bz$ and $\Bzbar$, 
the $\Bz\rto\pi^+\pi^-$ decay rate has a time-dependence given by 
\begin{eqnarray}
\label{eq:pipi-tdep}
{\cal P}_{\pi\pi}^q(\dt) = 
\frac{e^{-|\Delta{t}|/{\Tbz}}}{4{\Tbz}}
\left[1 + q\cdot 
\left\{ \spipi\sin(\dm\dt)   \right. \right. \nonumber \\
\left. \left.
   - \cpipi\cos(\dm\dt)
\right\}
\right],
\end{eqnarray}

where $\Tbz$ is the $\Bz$ lifetime, $\dm$ is the mass difference between the two
$\Bz$ mass eigenstates, $\dt = t_{CP}-t_{\rm tag}$, and the $b$-flavor charge
$q=+(-1)$ when the tagging $B$ meson is a $\Bz(\Bzbar$).
The \textit{CP}-violating asymmetry parameters $\spipi$ and $\cpipi ~(= -\apipi)$~\cite{bib:cpipi}
 defined 
in Eq.~(\ref{eq:pipi-tdep}) are expressed as
\begin{eqnarray}
\cpipi=\frac{1-|\lampipi|^2}{1+|\lampipi|^2}, \qquad
\spipi=\frac{2{\rm Im}\lampipi}{1+|\lampipi|^2},
\end{eqnarray}
where $\lampipi$ is a complex parameter that depends on both $\Bz$-$\Bzbar$ mixing and
on the amplitudes for $\Bz$ and $\Bzbar$ decay to $\pi^+\pi^-$.
A measurement of time-dependent \textit{CP}-violating asymmetries
in the mode $\Btopippim$ is sensitive to direct \textit{CP} violation and the
CKM angle $\alpha$.
If the decay proceeded only via a $b \rto u$ tree amplitude, 
$\spipi = \sin2\alpha$ and $\cpipi = 0$, or equivalently
$|\lambda_{\pi\pi}| = 1$.
The situation is complicated by the possibility of significant contributions
from gluonic $b \rto d$ penguin amplitudes that have a different weak phase
and additional strong phases.
In general, $\spipi$ is given by $\sqrt{1-\cpipi^2}\sin{2\alpha_{eff}}$
Here $\alpha_{eff}-\alpha$ (=$\theta$) depends on the magnitudes and
relative weak and strong phases of the tree and penguin amplitudes.
As a result, $\spipi$ may not be equal to $\sin2\alpha$ and direct \textit{CP}
violation, $\cpipi \neq 0$, may occur.

Candidate $B$ mesons are reconstructed kinematically using two variables,
the energy difference $\dE \equiv {\EBcms}-{\Ebeamcms}$   
and the beam-energy constrained mass $\Mbc
\equiv \sqrt{(\Ebeamcms)^2-(\pBcms)^2}$~\cite{bib:mES}, where
$\Ebeamcms$ is the cms beam energy, and $\EBcms$ and $\pBcms$ are the cms
energy and momentum of the $B$ candidate.

Charged tracks in $\Bz\rto h^+h^{\prime -}$ candidates are identified
 as charged pions or kaons.
Here $h$ and $h^\prime$ represent a $\pi$ or $K$. The
Belle Collaboration uses the likelihood ratio (KID) 
for a particle to be a $K^\pm$ meson,
which is based on the combined information from the Aerogel Cherenkov counter 
and CDC $dE/dx$ measurement.
Tracks are positively identified as pions with KID$<$0.4 for $\Btopippim$ candidates. 
The BaBar Collaboration uses the Cherenkov angle measurement $\theta_c$ 
from a detector of internally reflected Cherenkov light. 
The probability density function (PDF) from the difference between measured and 
expected values of $\theta_c$ is used in the extended likelihood function 
for the fit to extract yields and \textit{CP} parameters.

Background from the process $e^+e^- \rto \qq$ ~continuum~$(q=u,d,s,c)$ are suppressed by their 
event topology.
The Belle Collaboration forms signal and background likelihood functions ${\cal L}_S$
and ${\cal L}_{BG}$ from a Fisher discriminant determined from six modified
Fox-Wolfram moments~\cite{bib:Belle-Fisher} and 
the cms $B$ flight direction with respect to the beam axis. 
The continuum background is reduced by imposing requirements on the likelihood ratio
$LR$ = ${\cal L}_S/({\cal L}_S+{\cal L}_{BG})$ for different flavor-tagging dilution factor
intervals.
The BaBar Collaboration uses the angle $\theta_S$ between the sphericity axis of the $B$
candidate and the sphericity axis of the remaining particles in the cms frame, and
cut on $|\cos\theta_S|$.
The shapes of Fisher discriminant $\cal F$~\cite{bib:BaBar-Fisher} for
signal and background events are included as PDFs in the maximum likelihood fit.

Leptons, kaons, and charged pions that are not associated with the reconstructed
$B$ candidate are used to identify the flavor of the accompanying $B$ meson.

The vertex reconstruction algorithm is the same as that used for the $\sin2\beta$
( $\sin2\phi_1$ ) analysis.
The time difference $\dt$ is obtained from the measured distance
between the $z$ positions along the beam direction of the $\Bz_{\pi\pi}$
and $\Bz_{\rm tag}$ decay vertices and the boost factor $\beta\gamma$ of the
$e^+e^-$ system.

\begin{figure}[!htbp]
\begin{center}
\hbox to\hsize{\hss
\includegraphics[width=4.0cm]{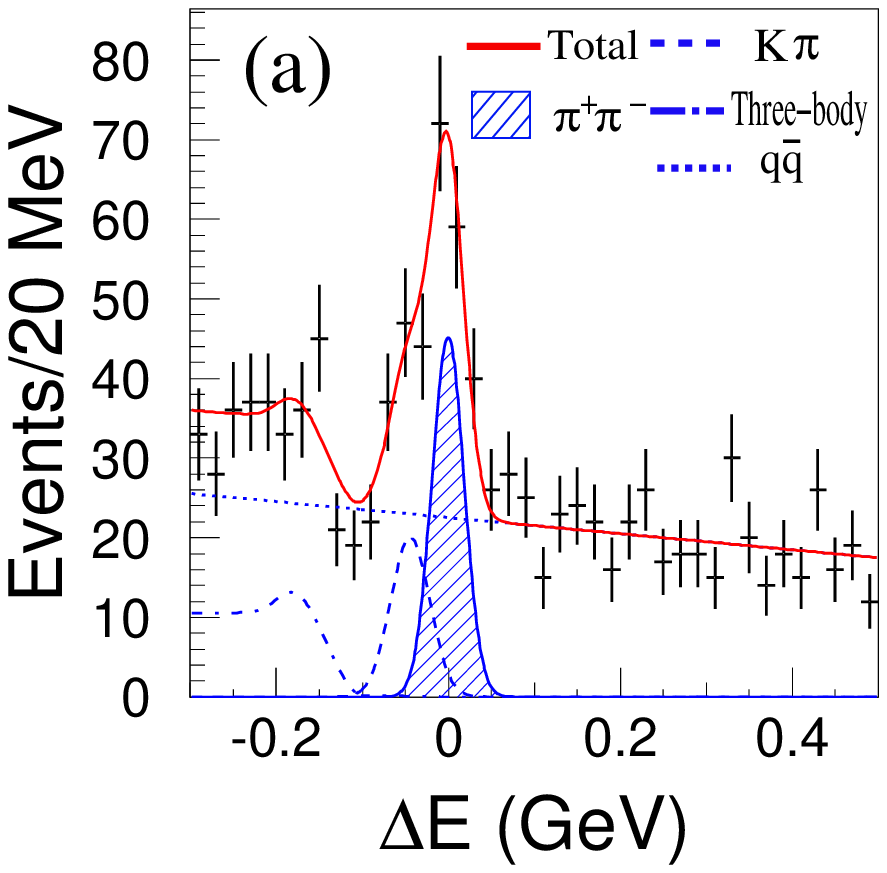}
\includegraphics[width=4.0cm]{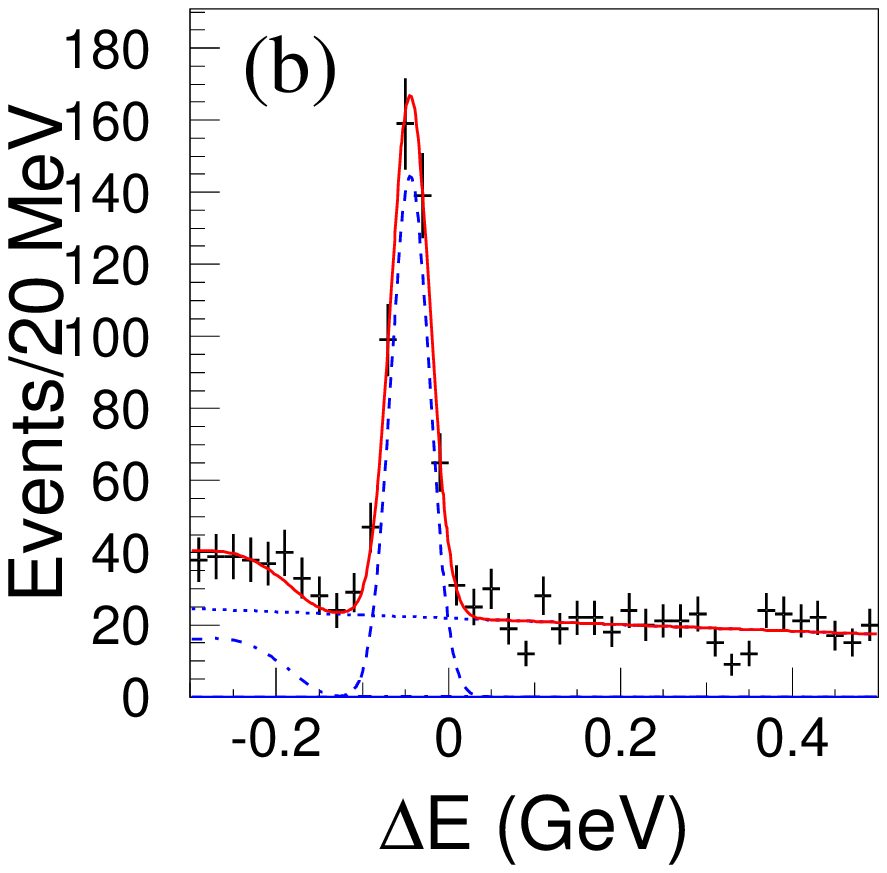}
\hss}
\end{center}
\caption{$\dE$ distributions in the $M_{bc}$ signal region 
for (a) $\Btopippim$ candidates and (b)
$\Bz \rto K^+\pi^-$ candidates with $\textit{LR} >0.825$
from the Belle Collaboration.
The sum of the signal and background functions is shown as a solid curve.
The solid curve with hatched area represents the $\pi^+\pi^-$ component,
the dashed curve represents the $K^+\pi^-$ component, the dotted curve represents
the continuum background, and the dot-dashed curve represents the charmless
three-body $B$ decay background component.
}
\label{fig:Belle-dE}
\end{figure}

\begin{figure}[!htbp]
\begin{center}
\hbox to\hsize{\hss
\includegraphics[width=4.0cm]{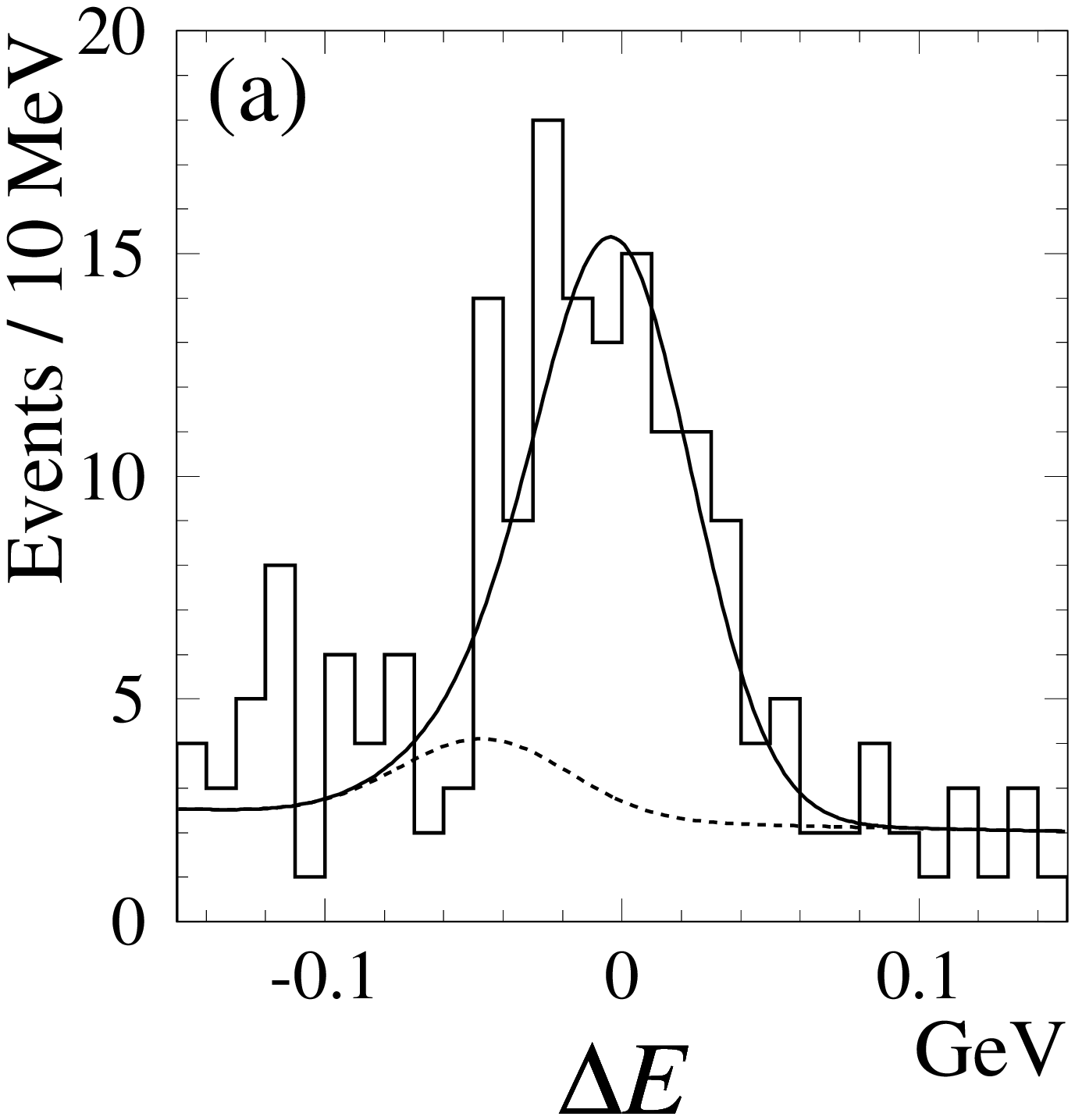}
\includegraphics[width=4.0cm]{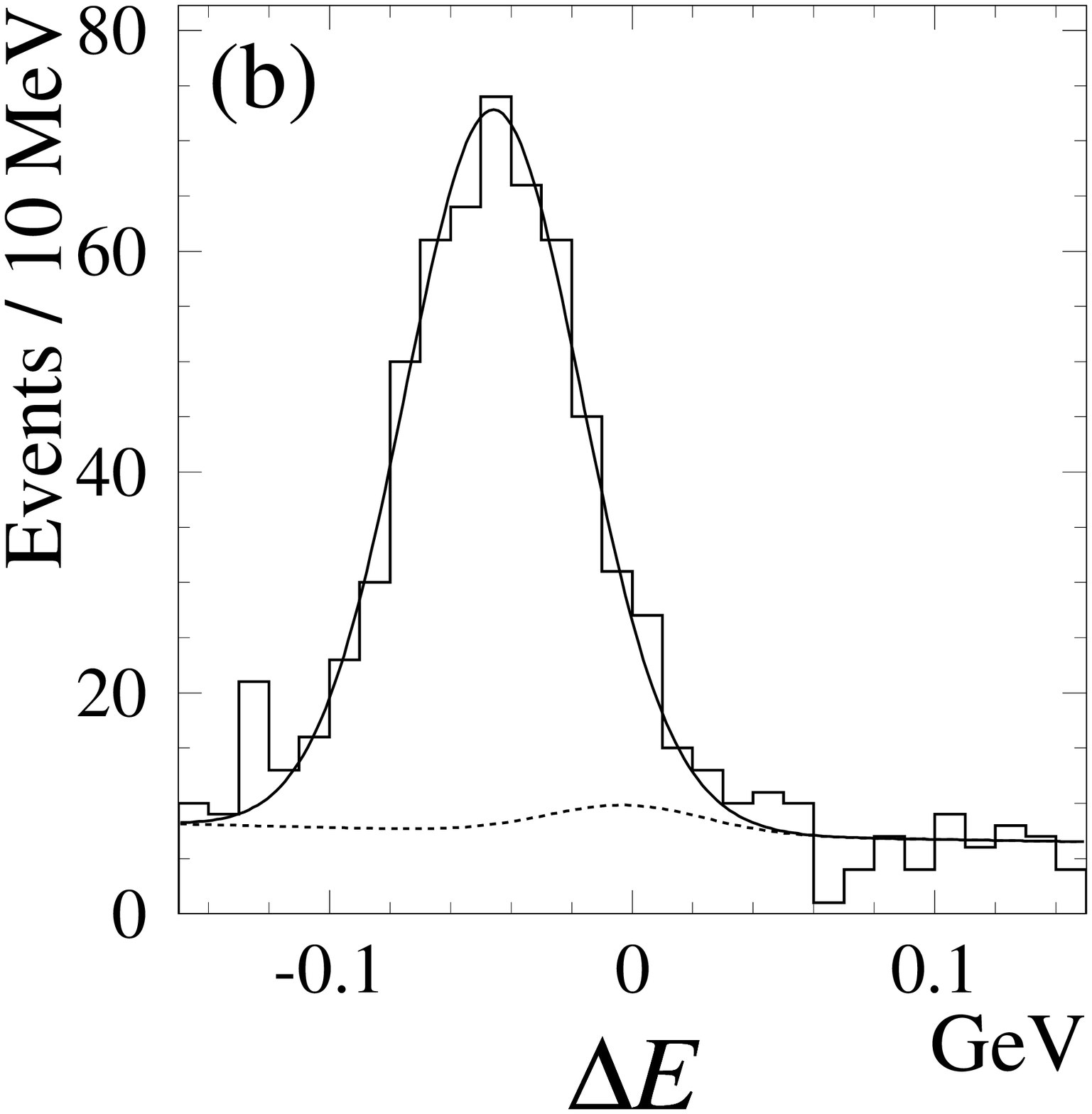}
\hss}
\end{center}
\caption{$\dE$ distributions for events enhanced in signal
(a) $\pi^+\pi^-$ and (b) $K^\mp\pi^\pm$ candidates
from the BaBar Collaboration. 
Solid curves represent projections of the maximum likelihood fit,
dashed curves represent $\qq$ and $\pi\pi \leftrightarrow K\pi$ cross-feed background.
}
\label{fig:BaBar-dE}
\end{figure}

\begin{figure}[!htbp]
\begin{center}
\hbox to\hsize{\hss
\includegraphics[width=\hsize]{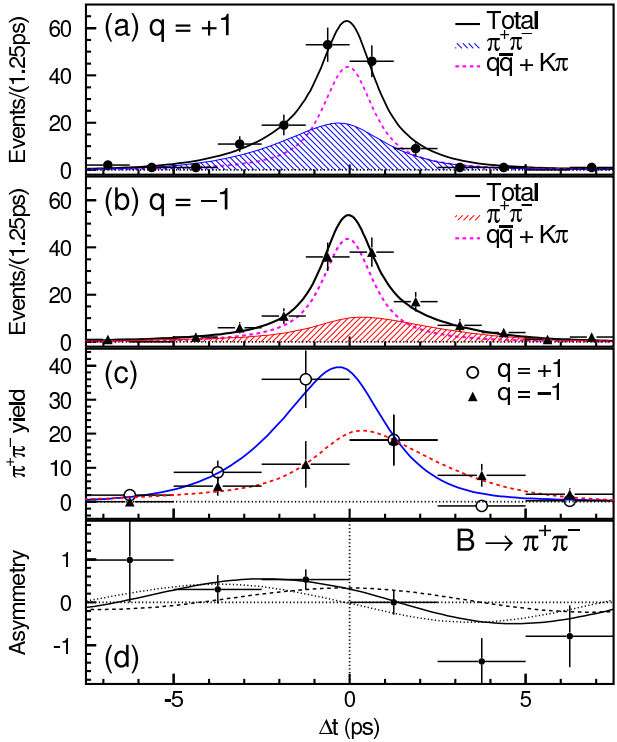}
\hss}
\end{center}
\caption{The raw, unweighted $\dt$ distributions for $\Btopippim$ candidates
with $\textit{LR} > 0.825$ in the signal region from the Belle Collaboration:
(a) candidates with $q = +1$, i.e. the tag side is identified as $\Bz$; 
(b) candidates with $q = -1$; 
(c) $\Btopippim$ yields after background subtraction;
(d) the \textit{CP} asymmetry for $\Btopippim$ after background subtraction.
In Figs. (a) through (c), the solid curves show the results of the unbinned maximum likelihood fit to the
$\dt$ distributions of the whole $\Btopippim$ candidates. In Fig. (d),
the dashed (dotted) curve is the contribution from the cosine (sine) term.}
\label{fig:Belle-asym}
\end{figure}

\begin{figure}[!htbp]
\begin{center}
\hbox to\hsize{\hss
\includegraphics[width=\hsize]{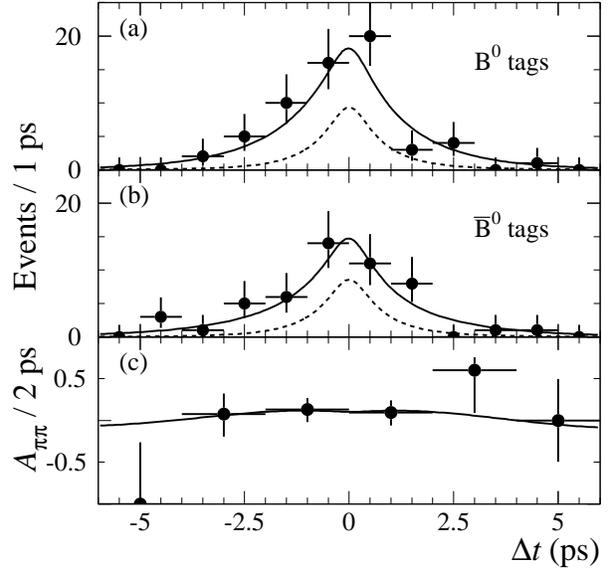}
\hss}
\end{center}
\caption{Distributions of $\dt$ for events enhanced in signal $\pi\pi$
decays from the BaBar Collaboration: 
the tag side is identified as (a) $\Bz$ or (b) $\Bzbar$,
and (c) the asymmetry as a function of $\dt$. Solid curves represent 
projections of the maximum likelihood fit, dashed curves represent
the sum of $\qq$ and $K\pi$ background events.}
\label{fig:BaBar-asym}
\end{figure}

Fig.~\ref{fig:Belle-dE} and Fig.~\ref{fig:BaBar-dE}
show distributions of $\dE$ for events enhanced
in signal $\pi^+\pi^-$ and $K^\mp\pi^\pm$ decays 
from the Belle Collaboration~\cite{bib:Belle-pipi} and
the BaBar Collaboration~\cite{bib:BaBar-pipi}, respectively.
The Belle and BaBar Collaborations obtained the following 
results using an unbinned maximum likelihood fit based on 85$\times 10^6$  and
88$\times 10^6 B{\Bbar}$ pairs, respectively,:
\begin{align}
\cpipi &= -0.77\pm{0.27}\pm{0.08}, &\spipi &= -1.23\pm{0.41}^{+0.08}_{-0.07} 
\nonumber \\
       & & &~~~~~~~~~~~~~~~~~~~~{\rm (Belle)},
\nonumber \\
\cpipi &= -0.30\pm{0.25}\pm{0.04}, &\spipi &= -0.02\pm{0.34}\pm{0.05} \nonumber \\
       & & &~~~~~~~~~~~~~~~~~~~~{\rm (BaBar)}.
\nonumber
\end{align}
The first and the second errors correspond to statistical and systematic errors,
respectively, unless otherwise stated.
The average values of $\cpipi$ and $\spipi$ are
\begin{eqnarray*}
\cpipi =-0.49\pm{0.19}, \qquad \spipi =-0.47\pm{0.26},
\end{eqnarray*}
and the difference of the results between the Belle and BaBar Collaborations 
is 2.2$\sigma$~\cite{bib:HFAG}.
In Fig.~\ref{fig:Belle-asym} and Fig.~\ref{fig:BaBar-asym} the $\dt$ distributions for events
enhanced in signal $\Bz\rto\pi^+\pi^-$ decays are shown for 
the Belle Collaboration~\cite{bib:Belle-pipi} and the
BaBar Collaborations~\cite{bib:BaBar-pipi}, respectively.

\begin{figure}[!htbp]
\begin{center}
\hbox to\hsize{\hss

\includegraphics[width=\hsize]{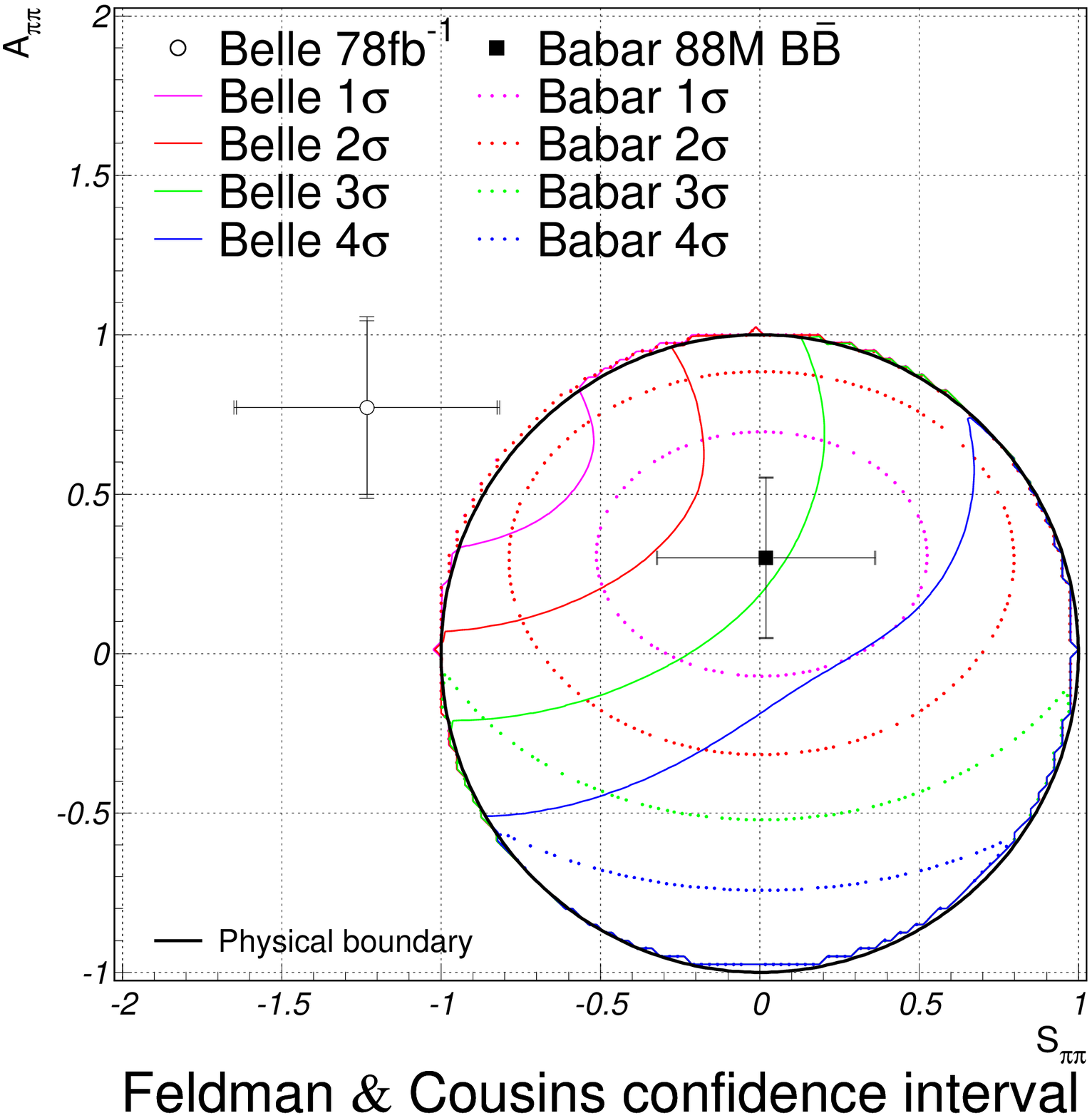}
\hss}
\end{center}
\caption{Confidence regions for $\apipi$ and $\spipi$ from the Belle 
and BaBar results.}
\label{fig:conf.int.}
\end{figure}

Fig.~\ref{fig:conf.int.} shows the two-dimensional confidence regions in
the $\apipi ( = -\cpipi )$ vs. $\spipi$ plane
using the Feldman-Cousins frequentist 
approach~\cite{bib:FC-frequentist}.
In order to form confidence intervals, the $\apipi$ and $\spipi$ 
distributions of the results of fits to MC pseudo-experiments for various input
values of $\apipi$ and $\spipi$ are used for the Belle result, and
the obtained errors of $\apipi$ and $\spipi$ are used for the BaBar result. 
The case that \textit{CP} symmetry is conserved, $\apipi = \spipi = 0$, is ruled out at the
99.93$\%$ confidence level (C.L.), equivalent to 3.4$\sigma$  significance for
Gaussian errors from the Belle result.
More data is necessary to clarify the difference between the Belle result and 
the BaBar result. 

\begin{figure}[!htbp]
\begin{center}
\hbox to\hsize{\hss
\includegraphics[width=\hsize]{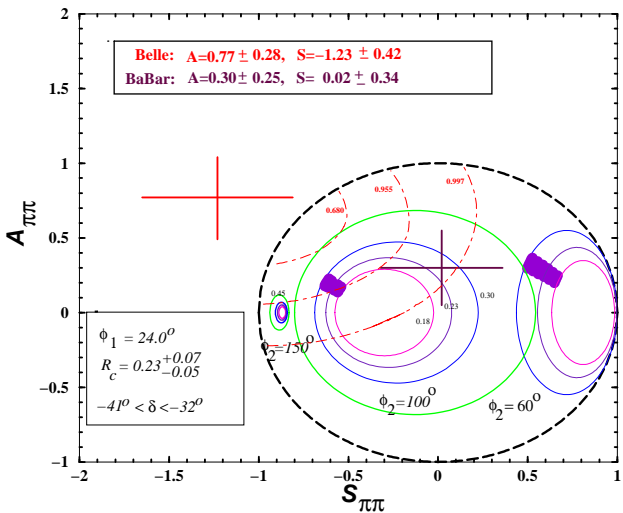}
\hss}
\end{center}
\caption{Plot of $\apipi$ versus $\spipi$ for various values of
$\phi_2$ with $\phi_1$=24.0$^\circ$, $0.18<R_C<0.30$, 
and $-41^\circ<\delta<-32^\circ$ in the pQCD method.
Here $R_C=|P/T|$.
Dark areas are allowed regions in the pQCD method for different $\phi_2$ values.
The results of $\apipi$ and $\spipi$ from the Belle 
and BaBar Collaborations and the confidence regions from the Belle Collabolation are
also shown.}
\label{fig:pQCD}
\end{figure}

\begin{figure}[!htbp]
\begin{center}
\hbox to\hsize{\hss
\includegraphics[width=\hsize]{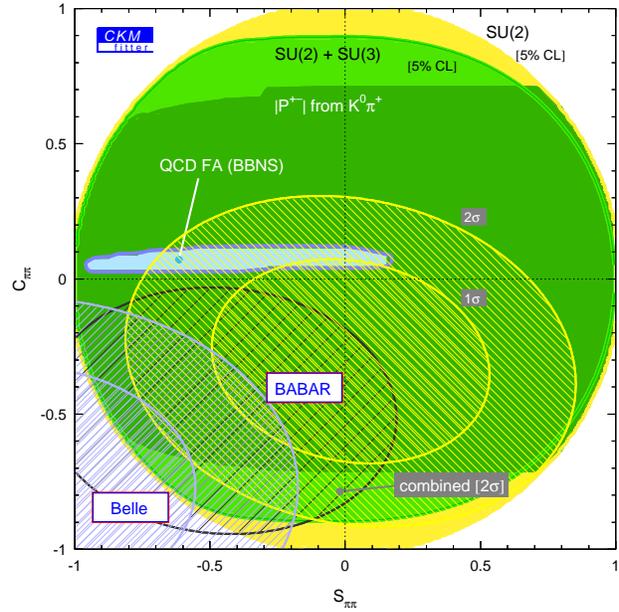}
\hss}
\end{center}
\caption{Predictions for $\cpipi ~( = -\apipi )$ and $\spipi$ 
for several analysis steps with experimental and theoretical constraints.
The Belle and BaBar results are shown.}
\label{fig:scpipi}
\end{figure}

\begin{figure}[!htbp]
\begin{center}
\hbox to\hsize{\hss
\includegraphics[width=\hsize]{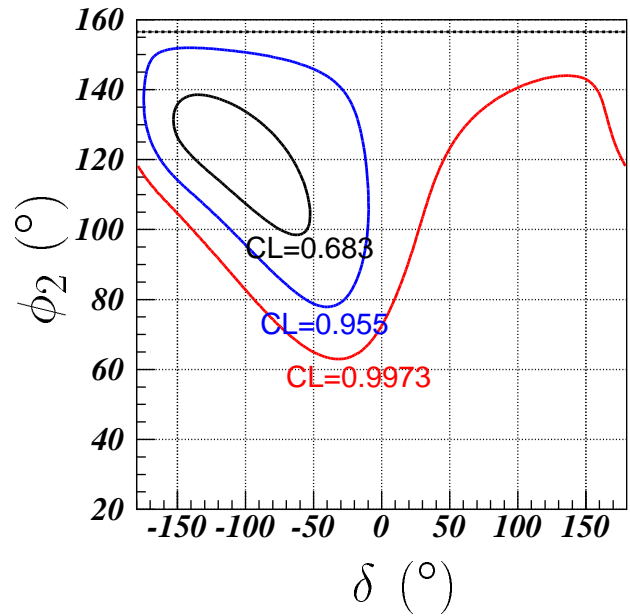}\hss}
\end{center}
\caption{The region for $\phi_2$ and $\delta$ which corresponds to 
the 68.3$\%$, 95.5$\%$, and 99.73$\%$ C.L. regions of 
$\apipi$ and $\spipi$ from the Belle result in Fig.~\ref{fig:pQCD}.
$\phi_1 = 23.5^\circ$ and $|P/T| = 0.45$.
The horizontal dashed line corresponds to $\phi_2 = 180^\circ - \phi_1$.}
\label{fig:Belle-phi2-delta}
\end{figure}

The decay amplitudes for $\Bz$ and $\Bzbar$ to $\pi^+\pi^-$ 
can be written by using the $c$-convention notation~\cite{bib:SA_TH_GR}:
\begin{eqnarray}
A(B^0\rightarrow\pi^+\pi^-) &=& -(|T|e^{i\delta_{T}}e^{i\phi_3}~~ +
|P|e^{i\delta_P}) , \nonumber \\
A(\bar{B}^0\rightarrow\pi^+\pi^-) &=& -(|T|e^{i\delta_{T}}e^{-i\phi_3} +
|P|e^{i\delta_P}) ,
\end{eqnarray}
where $T$ and $P$ are the amplitudes for the tree and penguin 
graphs and $\delta_T$ and $\delta_P$ are their strong phases. 
The expressions for $\spipi$ and $\apipi$ are
\begin{eqnarray}
\spipi &=& [ \sin 2\phi_2 + 2|P/T| \sin (\phi_1 - \phi_2) \cos \delta \nonumber \\
& & \mbox{} - |P/T|^2 \sin 2\phi_1]/\rpipi,  \nonumber \\
\apipi &=& - [2|P/T| \sin (\phi_2 + \phi_1) \sin \delta ]/\rpipi, \nonumber \\
\rpipi &=& 1 - 2|P/T| \cos \delta \cos (\phi_2 + \phi_1) + |P/T|^2 ,
\end{eqnarray}
where the strong phase difference $\delta$ $\equiv$ $\delta_P$ $-$ $\delta_T$.
When $\apipi$ is positive and $0^\circ<\phi_1+\phi_2<180^\circ$, $\delta$ is 
negative.
Fig.~\ref{fig:pQCD} shows the two-dimensional confidence regions in the
$\apipi$ vs. $\spipi$ plane together with the pQCD prediction~\cite{bib:pQCD}
for various values of $\phi_2 ~( = \alpha )$.
Fig.~\ref{fig:scpipi} shows predictions for $\cpipi ~( = -\apipi )$ and $\spipi$
for several analysis steps with experimental and theoretical 
constraints~\cite{bib:CKMfitter}.
In Fig.~\ref{fig:Belle-phi2-delta}, the interpretation
of the confidence regions of $\apipi$ vs. $\spipi$ is shown 
in terms of $\phi_2$ and $\delta$ 
for the Belle data~\cite{bib:Belle-pipi}. 
The range of $\phi_2$ that corresponds to the 95.5$\%$ C.L. region of $\apipi$
and $\spipi$ is $78^\circ\leq\phi_2\leq152^\circ$ for 
$\phi_1=23.5^\circ$~\cite{bib:phi1} and
$0.15\leq|P/T|\leq0.45$~\cite{bib:POT}.
The result is in agreement with constraints on the unitarity triangle from other 
measurements~\cite{bib:Nir}. 
Other interpretations for the current results 
can be found in ref.~\cite{bib:CKMfitter}.

Using isospin relations~\cite{bib:GLSS}, we can constrain the difference, $\theta$ between
$\alpha_{eff}$ and $\alpha$. From the central values of the
recent world average values of the
branching ratios of $\Bz\rto\pi^+\pi^-$, $B^+\rto\pi^+\pi^0$ and the 
90$\%$ C.L. upper limit on the $\Bz\rto\pi^0\pi^0$ branching ratio~\cite{bib:HFAG-rare}
 together with $\cpipi$,
the upper limit on $\theta$ is 54$^\circ$.
 
\section{$\Bz\rto\rho\pi\rto\pi^+\pi^-\pi^0$ decays}

In principle, the CKM angle $\alpha$ can be measured in the presence of  penguin
contributions using a full Dalitz plot analysis of the final state.
However, there are difficulties of combinatorics and lower efficiency in three-body
topology with $\pi^0$ and large backgrounds from misreconstructed signal events and
other decays. In order to extract $\alpha$ cleanly, data with large statistics are
required.

Unlike $\Bz\rto\pi^+\pi^-$ decay, $\Bz\rto\rho^\pm\pi^\mp$ decay is not a \textit{CP}
eigenstate, and four flavor-charge configurations 
($\Bz(\Bzbar)\rto\rho^\pm\pi^\mp$) must be considered.
Following a quasi-two-body approach~\cite{bib:BaBar-Phys-Book},
the analysis is restricted to the two regions of the $\pi^\pm\pi^0{h^\pm}$
Dalitz plot ($h$ = $\pi$ or $K$) that are dominated by $\rho^\pm{h^\mp}$.
The decay rate is given by
\begin{align}
&f^{\rho^{\pm}h^{\mp}}_{q}(\dt)
=(1\pm\ArhCP)\frac{e^{-|\Delta t|/{\Tbz}}}{4{\Tbz}} \nonumber \\
&~~~~~~~~~~~~~~~\times[1+q{\cdot}((\Srh\pm\Delta\Srh) \sin (\dm\dt) \nonumber \\
&~~~~~~~~~~~~~~~~~~~~~~-q{\cdot}(\Crh\pm\Delta\Crh) \cos (\dm\dt))],
\end{align}
where $\Delta t$ = $t_{\rho h}$ $-$ $t_{\rm tag}$ as the time interval 
between the decay of $B^0_{\rho h}$ and that of the other $B^0$ meson in the event,
$B^0_{\rm tag}$.

\begin{figure}[!htbp]
\begin{center}
\hbox to\hsize{\hss
\includegraphics[width=\hsize,angle=-90]{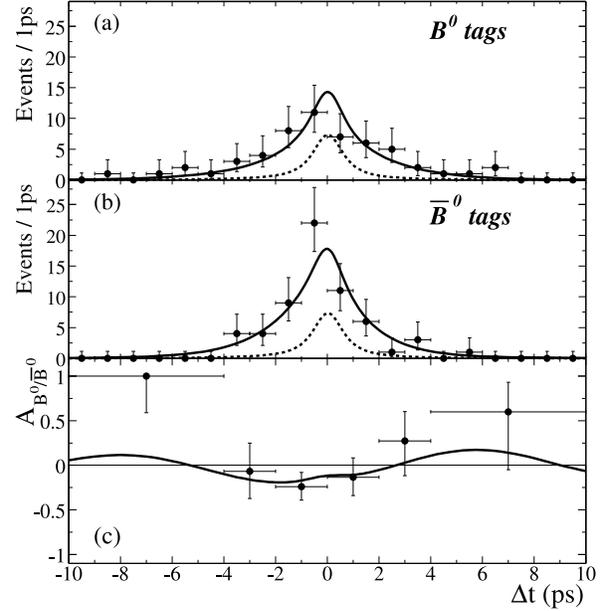}
\hss}
\end{center}
\caption{
Time distributions for events selected to enhance the $\rho\pi$ signal tagged as
(a) $\Bz$-tag and (b) $\Bzbar$-tag, and (c) time-dependent asymmetry
between $\Bz$-tag and $\Bzbar$-tag from the BaBar Collaboration~\cite{bib:BaBar-rhopi}.
The solid curve is a likelihood projection of the fit result.
The dashed line is the sum of $B$- and continuum-background contributions.
}
\label{fig:BaBar-rhopi}
\end{figure}

The time- and flavor-integrated charge asymmetries $\ArpCP$ and $\ArKCP$ measure
direct \textit{CP} violation.
For the $\rho\pi$ mode, the quantities $\Srp$ and $\Crp$ parameterize 
mixing-induced \textit{CP} violation related to the CKM angle $\alpha$, and
flavor-dependent direct \textit{CP} violation, respectively.
$\Delta\Crp$ describes the asymmetry between the rates
$\Gamma (\Bz\rto\rho^+\pi^-) + \Gamma (\Bzbar\rto\rho^-\pi^+)$ and
$\Gamma (\Bz\rto\rho^-\pi^+) + \Gamma (\Bzbar\rto\rho^+\pi^-)$.
$\Delta\Srp$ is related to the strong phase difference between the
amplitudes contributing to $\Bz\rto\rho\pi$ decays.
One finds the relations $\Srp\pm\Delta\Srp$ = 
$\sqrt{1-(\Crp\pm\Delta\Crp)^2}\sin(2\alpha^\pm_{\rm eff}\pm\delta)$,
where 2$\alpha^\pm_{\rm eff}$ = 
arg[$(q/p)({\bar{A}^\pm_{\rho\pi}}/{A^\mp_{\rho\pi}})$],
$\delta$ = arg[$A^-_{\rho\pi}/A^+_{\rho\pi}$],
arg[q/p] is the $\Bz$-$\Bzbar$ mixing phase, and
$A^+_{\rho\pi}(\bar{A}^+_{\rho\pi})$ and
$A^-_{\rho\pi}(\bar{A}^-_{\rho\pi})$ are the transition amplitudes of the
processes $\Bz(\Bzbar)\rto\rho^+\pi^-$ and
$\Bz(\Bzbar)\rto\rho^-\pi^+$, respectively.
The angles $\alpha^\pm_{\rm eff}$ are equal to $\alpha$ 
if contributions from penguin amplitudes are absent.
For the self-tagging $\rho K$ mode, the values of the four time-dependent 
parameters are $\CrK = 0$, $\dCrK = -1$, $\SrK = 0$, and $\dSrK = 0$.
The results on direct \textit{CP} violation can be expressed using the asymmetries
\begin{eqnarray}
\Apm &=& \frac{N(\Bzbarrp \rto \rppm)-N(\Bzrp \rto \rmpp)}
      {N(\Bzbarrp \rto \rppm)+N(\Bzrp \rto \rmpp)} \nonumber \\ 
&=& \frac{\ArpCP-\Crp-\ArpCP\cdot{\DCRP}}{1-{\DCRP}-\ArpCP\cdot\Crp}\\
\Amp &=& \frac{N(\Bzbarrp \rto \rmpp)-N(\Bzrp \rto \rppm)} 
      {N(\Bzbarrp \rto \rmpp)+N(\Bzrp \rto \rppm)} \nonumber \\
&=& \frac{\ArpCP+\Crp+\ArpCP\cdot{\DCRP}}{1+{\DCRP}+\ArpCP\cdot\Crp}
\end{eqnarray}

With a data sample of 89 million $B\Bbar$ pairs~\cite{bib:BaBar-rhopi}, 
the BaBar Collaboration found
428$^{+34}_{-33}$(stat) $\rho\pi$
and 120$^{+21}_{-20}$(stat) $\rho K$ events 
and 
the following measurements of the \textit{CP} violation parameters are obtained:
\begin{eqnarray*}
\ArpCP &=& -0.18\pm{0.08}\pm{0.03},\\
\Crp   &=& +0.36\pm{0.18}\pm{0.04},\\
\Srp   &=& +0.19\pm{0.24}\pm{0.03}.
\end{eqnarray*}
For the other parameters in the description of the 
$B^0(\Bzbar) \rto \rho\pi$ decay-time dependence,
\begin{eqnarray*}
\DCRP  &=& +0.28\pm{0.19}\pm{0.04},\\
\DSRP  &=& +0.15\pm{0.25}\pm{0.03},
\end{eqnarray*}
are found.
For the asymmetries $\Apm$ and $\Amp$, which probe direct \textit{CP} violation,
\begin{eqnarray*}
\Apm &=& -0.62^{+0.24}_{-0.28}\pm 0.06,\\
\Amp &=& -0.11^{+0.16}_{-0.17}\pm{0.04},
\end{eqnarray*}
are measured.
The raw time-dependent asymmetry in the tagging categories dominated by
kaons and leptons is shown in Fig.~\ref{fig:BaBar-rhopi}.

\section{Prospects}

Table~\ref{tab:future} shows the expected error on \textit{CP}-violating
parameters in $\Btopippim$ and $\Bz\rto\rho^\pm\pi^\mp$ decays with accumulated
luminosities of 140 fb$^{-1}$, 
400 fb$^{-1}$, 3000 fb$^{-1}$ (3 ab$^{-1}$), and 30000 fb$^{-1}$ (30 ab$^{-1}$)
in the future.

\begin{figure}[!htbp]
\begin{center}
\hbox to\hsize{\hss
\includegraphics[width=\hsize]{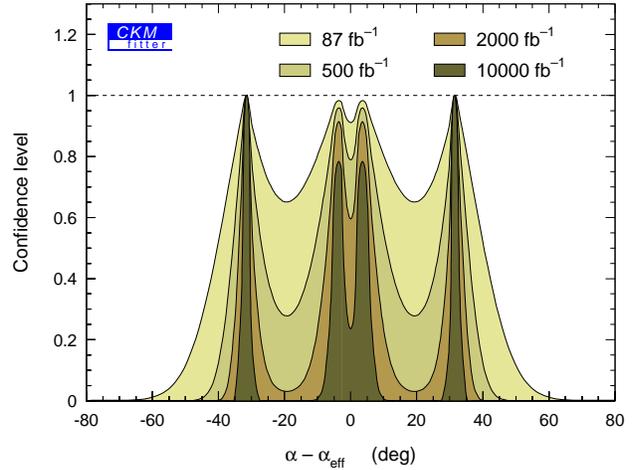}
\hss}
\end{center}
\caption{
$\alpha_{eff}-\alpha$ 
for the ICHEP$^\prime$02 central values of the $B\rto\pi\pi$ branching fractions
and the central values of $\spipi$ and $\cpipi$ from the BaBar measurement at
luminosities of the current and future B-factories
( 87 fb$^{-1}$, 500 fb$^{-1}$, 2~ab$^{-1}$, and
10~ab$^{-1}$ ).
}
\label{fig:pipi-future}
\end{figure}

Fig.~\ref{fig:pipi-future} shows the prospects of $\alpha_{eff}-\alpha$ 
for the ICHEP$^\prime$02 central values of the $B\rto\pi\pi$ branching ratios and
the central values of $\spipi$ and $\cpipi$ from the BaBar measurement at
luminosities of the current and future B-factories
( 87~fb$^{-1}$, 500~fb$^{-1}$, 2~ab$^{-1}$, 10~ab$^{-1}$ ).
The inner and outer boarders can be obtained from the isospin analysis
when $\Bz\rto\pi^0\pi^0$ flavors are tagged and not tagged, respectively.
Only a luminosity of around 10~ab$^{-1}$
allows to separate the solutions.

For $B\rto\rho\pi$ decays, the isospin analysis 
is not feasible yet with the present statistics of the $B$ factories. 
In~\cite{bib:CKMalpha-rhopi}, the projections into
the future full SU(2) analysis was demonstrated.
If the branching fraction of $\Bz\rto\rho^0\pi^0$ is below the experimental 
sensitivity, a strong constraint on $\alpha$ is expected above luminosity of
around 2~ab$^{-1}$.
In this workshop, theoretical problems such as form factors and 
$\sigma$ meson~\cite{bib:sigma-meson},
and experimental problems for several sources of backgrounds were pointed out.

Detailed discussions can be found in ~\cite{bib:CKMalpha-pipi} for
$\Btopippim$ and in ~\cite{bib:CKMalpha-rhopi} for
$\Btorp$ at this workshop.

\begin{table}[!htbp]
\begin{center}
\begin{tabular}{|c|c|c|c|c|}\hline
parameters & 140 fb$^{-1}$ & 400 fb$^{-1}$ & 3 ab$^{-1}$ & 30 ab$^{-1}$ \\ \hline
$\apipi$ & 0.21 & 0.13 &  0.05 &  0.02  \\
$\spipi$ & 0.31 & 0.19 &  0.07 &  0.03  \\ \hline
$\ArpCP$ & 0.07 & 0.04 &  0.02 &  0.008 \\
$\Crp$   & 0.14 & 0.09 &  0.03 &  0.013 \\
$\Srp$   & 0.18 & 0.11 &  0.04 &  0.014 \\ \hline
\end{tabular}
\end{center}
\caption{The
errors of \textit{CP}-violating parameters in 
$\Btopippim$ and $\Bz\rto\rho^\pm\pi^\mp$ decays at 140~fb$^{-1}$, 400~fb$^{-1}$, 
3~ab$^{-1}$, and 30~ab$^{-1}$, assuming
that statistical and systematic errors are proportional to
1/$\sqrt{\cal L}$ and 1/$\sqrt[4]{\cal L}$, respectively.
Here {\cal L} is accumulated luminosity. }
\label{tab:future}
\end{table} 

\section{Summary}
In summary, 
the Belle and BaBar Collaborations obtain the following measurements of the
\textit{CP}-violating asymmetry parameters in $\Bz\rto\pi^+\pi^-$ decays:
\begin{align}
\apipi &= +0.77\pm{0.27}\pm{0.08}, &\spipi &= -1.23\pm{0.41}^{+0.08}_{-0.07} 
\nonumber \\
       & & &~~~~~~~~~~~~~~~~~~~~{\rm (Belle)}, \nonumber \\
\apipi &= +0.30\pm{0.25}\pm{0.04}, &\spipi &= -0.02\pm{0.34}\pm{0.05}
\nonumber \\
       & & &~~~~~~~~~~~~~~~~~~~~{\rm (BaBar)}, 
\nonumber
\end{align}

The following measurements of the \textit{CP}-violating asymmetry parameters in
$\Bz\rto\rho\pi$ decays using a quasi two-body analysis 
are obtained by the BaBar Collaboration:
\begin{align}
\ArpCP &= -0.18\pm{0.08}\pm{0.03}, \nonumber \\
\Crp   &= +0.36\pm{0.18}\pm{0.04}, & \DCRP  &= +0.28\pm{0.19}\pm{0.04}, \nonumber \\
\Srp   &= +0.19\pm{0.24}\pm{0.03}, & \DSRP  &= +0.15\pm{0.25}\pm{0.03}. \nonumber
\end{align}
For the asymmetries $\Apm$ and $\Amp$, which probe direct $CP$ violation,
\begin{eqnarray*}
\Apm = -0.62^{+0.24}_{-0.28}\pm 0.06, \qquad
\Amp = -0.11^{0.16}_{-0.17}\pm{0.04},
\end{eqnarray*}
are obtained.






 




\begin{thebibliography}{9}
\bibitem{bib:CC}
Throughout this paper, the inclusion of the charge conjugate mode decay is implied
unless otherwise stated.
\bibitem{bib:BaBar-Fisher}
${\cal F} = 0.53 - 0.60 \times \sum_i{p^*_i} +
1.27\times\sum_i{p^*_i}|\cos(\theta^*_i)|^2$, where $p^*_i$ is the momentum of
particle $i$ and $\theta^*_i$ is the angle between its momentum and the $B$ thrust
axis in the cms frame.
The sum is over all particles in the event excluding
those of the $B$ candidate.
\bibitem{bib:sigma-meson}
A.D.~Polosa, these proceedings.
The $\sigma$ meson was introduced as a very broad light resonance.
The E791 Collaboration had an evidence of a very broad scalar resonance
with
$m_\sigma = 478 {\rm MeV}$ and
$\Gamma_\sigma = 324 {\rm MeV}$~\cite{bib:E791}.
$\Bz\rto\sigma\pi^0$ can spoil the penguin relation on 
the isospin analysis in $\Bz\rto\rho\pi$.
\end{thebibliography}
\end{document}